\documentclass[onecolumn,nofootinbib,preprintnumbers,amsmath,amssymb,longbibliography,10pt]{revtex4}
\usepackage{array,multirow,graphicx}
\usepackage{dcolumn}
\usepackage{txfonts}
\usepackage{newlfont}
\usepackage{times}
\usepackage{bm}
\usepackage[colorlinks,citecolor=blue,urlcolor=blue,linkcolor=blue]{hyperref}
\usepackage[figtopcap]{subfigure}
\usepackage{color,soul}
\soulregister\cite7
\soulregister\ref7
\soulregister\pageref7
\begin{document}
\preprint{FU-PCG-45}
\title{ Charged rotating black holes coupled with nonlinear electrodynamics Maxwell field in the mimetic gravity}
\author{G.G.L. Nashed$^{1,2,3}$}%
\email{nashed@bue.edu.eg}
\author{W. El Hanafy$^{1,2}$}%
\email{waleed.elhanafy@bue.edu.eg}
\author{Kazuharu Bamba$^{4}$}%
\email{bamba@sss.fukushima-u.ac.jp}
\affiliation{$^{1}$Centre for Theoretical Physics, The British University in Egypt, P.O. Box 43, El Sherouk City, Cairo 11837, Egypt}
\affiliation{$^{2}$Egyptian Relativity Group (ERG), Cairo University, Giza 12613, Egypt}
\affiliation{$^{3}$Mathematics Department, Faculty of Science, Ain Shams University, Cairo 11566, Egypt}
\affiliation{$^{4}$Division of Human Support System, Faculty of Symbiotic Systems Science, Fukushima University, Fukushima 960-1296, Japan}

\begin{abstract}
In mimetic gravity, we derive $D$-dimension charged black hole solutions having flat or cylindrical horizons with zero curvature boundary. The asymptotic behaviours of these black holes behave as (A)dS. We study both linear and nonlinear forms of the Maxwell field equations in two separate contexts. For the nonlinear case, we derive a new solution having a metric with monopole, dipole and quadrupole terms. The most interesting feature of this black hole is that its dipole and quadruple terms are related by a constant. However, the solution reduces to the linear case of the Maxwell field equations when this constant acquires a null value. Also, we apply a coordinate transformation and derive rotating black hole solutions (for both linear and nonlinear cases).  We show that the nonlinear black hole has stronger curvature singularities than the corresponding known black hole solutions in general relativity. We show that the obtained solutions could have at most two horizons. We determine the critical mass of the degenerate horizon at which the two horizons coincide. We study the thermodynamical stability of the solutions. We note that the nonlinear electrodynamics contributes to process a second-order phase transition whereas the heat capacity has an infinite discontinuity.

\end{abstract}
\maketitle
\section{Introduction}\label{S1}
In the last few decades, several cosmological observations have confirmed that our universe is dominated by dark components: dark matter and dark energy \cite{Riess:1998cb,Perlmutter:1998np,Ade:2015xua}. The competition to criticize the origin and the nature of these ingredients either within observational or theoretical framework is going on equal footing. In the observational framework, it has been shown that about 26\% of the energy content in our universe belongs to the sector of dark matter, however, about 69\% specifies the dark energy. In the theoretical framework, modified gravity theories are the most appealing by invoking a geometrical origin to explain these ingredients \cite{Capozziello:2002rd,Carroll:2003wy,Nojiri:2003rz,Dolgov:2003px,Chiba:2003ir,
Carroll:2004de,Nojiri:2006ri,Woodard:2006nt,Nojiri:2006be,Wanas:2012ve,Clifton:2011jh,Capozziello:2011et,Bamba:2012cp}. Among modified gravity theories, we particularly mention two approaches that gain attention in literature, the curvature based $f(R)$ gravity, c.f.~ \cite{Nojiri:2003ft,Capozziello:2005ku,Amendola:2006kh,Capozziello:2006dj,Cembranos:2008gj,Nojiri:2010wj} and the torsion based $f(T)$ teleparallel gravity, c.f.~ \cite{Boehmer:2011gw,Dong:2012en,Nashed:2015pda,Bamba:2016gbu,Paliathanasis:2014iva,Nashed:2016tbj,Awad:2017ign,Awad:2017yod} (for recent reviews on modified theories of gravity and the issue of dark energy, see, for example~\cite{Nojiri:2010wj, Capozziello:2011et, Capozziello:2010zz, Bamba:2015uma, Cai:2015emx, Nojiri:2017ncd, Bamba:2012cp}). Another modified scenario that has been proposed recently is the mimetic gravity one \cite{Chamseddine:2013kea}. Mimetic theory has many applications in  cosmology  \cite{Chamseddine:2016ktu,Chamseddine:2016uef,Saadi:2014jfa,Haghani:2015iva,Baffou:2017pao} as well as in the solar system \cite{Myrzakulov:2015sea,Addazi:2017cim,Astashenok:2015qzw,Myrzakulov:2015kda,Babichev:2016jzg}, see also the recent review \cite{Sebastiani:2016ras}.\\

Although this theory provides geometrical foundations to explain dark components as in modified gravity, it presents a distinguishable framework to deal with that problem. The construction of the mimetic theory can be derived from the general relativity theory by separating the conformal degree of freedom of the gravitational field via re-parameterizing the physical metric by an auxiliary metric with a mimetic field $\phi$. The equations of motion in the mimetic theory are characterized by an extra term sourced by the mimetic field. The mimetic scalar field is generated by a singular limit of the general informal transformation which is not invertible \cite{Deruelle:2014zza,Yuan:2015tta,Domenech:2015tca}, and its kinetic term is provided to satisfy the kinematical constraint
\begin{equation}\label{con}
g^{\mu\nu}\partial_\mu \phi \, \partial_\nu \phi=-1.
\end{equation}

Using the above constraint, the equations of motion have shown to be traceless, where the scalar field mimics pressureless cold dark matter (CDM) just as in general relativity even for vacuum solution \cite{Sebastiani:2016ras}. While generalizing the model by invoking potential functions, one can produce a unified cosmic history nothing to speak of bouncing cosmology \cite{Nojiri:2014zqa,Odintsov:2016imq,Oikonomou:2016fxb,Dutta:2017fjw}. The mimetic gravity has shown to be invariant under Weyl transformation \cite{Firouzjahi:2017txv,Hammer:2015pcx}. On its perturbation level, the sound speed of the scalar fluctuations is exactly zero, $c_s=0$, so to obtain a successful inflation one may impose a higher-derivative term $(\Box \phi)^2$ to grant the scalar fluctuation a nonzero sound speed \cite{Chamseddine:2014vna,Mirzagholi:2014ifa}. This feature may have important impacts on large structure and galaxy formation specially for very small but not vanishing values of the sound speed. In particular, the higher-derivative term departs the properties of the mimetic dark matter from being a perfect dust-like fluid \cite{Chaichian:2014qba,Malaeb:2014vua,Ali:2015ftw}. In these cases, imperfect dark matter could be relevant to the missing-satellites problem and the core-cusp problem \cite{Capela:2014xta}.\\

One of the crucial issues in mimetic gravity is  stability. In fact, there are many studies on the stability of mimetic gravity . Although, the theory does not suffer from gradient instabilities or those are associated to higher derivatives (Ostrogradski ghost), the theory might have ghost instability \cite{Achour:2016rkg}. However, the later can be resolved by coupling higher derivatives of the mimetic field to curvature. Besides that it has been shown that the strong coupling scale can be raised to 10 TeV \cite{Ramazanov:2016xhp}. This cutoff scale on spatial momenta of ghosts makes the mimetic matter scenario is phenomenologically viable. At a lower scale, it reduces to GR supplemented by a fluid with small positive sound speed $c_s^2 \leq 10^{-20}$, which is compatible with the observations of the photon flux from vacuum decay $c_s^2 \geq 10^{-42}$. Also the authors of \cite{Barvinsky:2013mea, Chaichian:2014qba}, independently, found that the theory is stable if the density of the mimetic field is positive. This preliminary analysis favors the solutions with de Sitter backgrounds in presence of a cosmological constant, since in this case both the curvature and the trace of the matter energy-momentum tensor contribute to hold the energy density positive. However, the theory might still suffer from the gravitational instability associated with caustic surfaces of the geodesic flow \cite{Barvinsky:2013mea}. Another work using the effective theory approach has been developed to overcome the problem of gradient instability \cite{Hirano:2017zox}. Also, it has been shown that the mimetic constraint, namely Eq. (\ref{con}), can be implemented in the action using a Lagrangian multiplier \cite{Nojiri:2017ygt}. Using a suitable gravitational Lagrange multiplier could provide a healthy mimetic gravity free from ghosts \cite{Nojiri:2017ncd}. Moreover, in Gauss-Bonnet theories ghost modes can be removed at the level of equations of motion \cite{Nojiri:2018ouv}.\\


Recently one can find several variants of mimetic gravity: The mimetic $f(R)$ gravity, where some quantum/string corrections can be taken into account by adding higher-order curvature invariants to the action of mimetic gravitation theory \cite{Nojiri:2014zqa,Leon:2014yua}. Same technique has been applied to construct mimetic Gauss-Bonnet theories $f(G)$ \cite{Astashenok:2015haa}, mimetic $f(R, T)$ \cite{Momeni:2015gka}, mimetic $f(R, \phi)$ \cite{Mirzagholi:2014ifa}, mimetic covariant Horava-like gravity \cite{Myrzakulov:2015nqa,Cognola:2016gjy}, mimetic Galileon gravity \cite{Haghani:2014ita,Rabochaya:2015haa}, mimetic Horndeski gravity \cite{Arroja:2015wpa}, unimodular-mimetic $f(R)$ gravity \cite{Odintsov:2016imq}, mimetic Born-Infeld gravity \cite{Bouhmadi-Lopez:2017lbx,Chen:2017ify} and non-local mimetic $f(R)$ gravity \cite{Myrzakulov:2016hrx}. Notably, the importance  of higher-derivative invariants in the Lagrangian of the mimetic theory  has been analyzed in \cite{Ramazanov:2015pha,Paston:2017das,Hirano:2017zox,Zheng:2017qfs,Cai:2017dxl,Takahashi:2017pje,Gorji:2017cai}. There are many other amendments of the mimetic theory have been considered, for example by including the  vector-tensor mimetic gravity \cite{Kimura:2016rzw}, bi-scalar mimetic models \cite{Saridakis:2016mjd}, the one in which the implementation of the limiting curvature hypothesis is considered to solve the issues of the  cosmological singularity \cite{Chamseddine:2016uef,Chamseddine:2016ktu} and the braneworld mimetic gravity \cite{Sadeghnezhad:2017hmr}. Also, the models where the mimetic field couples to matter non-minimally \cite{Vagnozzi:2017ilo} and the currents of baryon number have been also discussed in \cite{Shen:2017rya}.\\

Modified gravity theories are usually tested via cosmological models and black hole physics. The aim of this work is to test mimetic gravity coupled to linear/nonlinear electrodynamics in the black hole physics domain. According to ``no hair" theorem, the black hole is characterized by three conserved quantities: ADM mass, spin and charge. Static spherically symmetric spacetime is known as Schwarzschild black hole, rotating case is known as Kerr black hole, when static spherically symmetric black hole gains charges, it is called Reissner-Nordstr\"{o}m (RN) and the rotating charged black hole is known as Kerr-Newmann (KN). The later type of black holes, interestingly, undergoes a phase transition associated with an infinite discontinuity of the black hole heat capacity \cite{Davies:1978mf}. In practice, mass and spin are tested and confirmed by observations. However, charged black holes are believed to be not existing in real scenarios. This is because  that requires a rapid neutralization. On the other hand, it is interesting to investigate mechanisms of producing charged black holes. This has been examined even within classical framework, c.f \cite{Wald:1974np}, others are to study the charge production mechanism by involving dark matter \cite{DeRujula:1989fe,Sigurdson:2004zp,Cardoso:2016olt}. Charged black hole solutions have been examined in several theoretical aspects c.f.~\cite{Nashed:2005kn,Nashed:2007cu,Nashed:2008ys,Nashed:2009hn,Nashed:2004pn}. However, the mimetic gravity has been proven to be a good candidate to describe dark matter. In this sense, we find studying charged black hole solutions in mimetic gravity as in the present paper is a step for more investigation in this regard. In astrophysics domain, the effect of charge on the merge rate of binary black holes has been studied recently \cite{Zhang:2016rli,Zhu:2018tzi}. In \cite{Zhang:2016rli}, it has been shown that a fast radio burst (FRB) or a gamma-ray burst (GRB) can be explained, depending on the value of the black hole charge. It sets lower limits on the charge necessary to produce each phenomenon. It has been shown that, for a 10 $M_\odot$ black hole, the merger can produce a FRB, if the charge of one members of the black hole charge is more than $\sim 10^{12}$ Coulombs. If its charge is more than $\sim 10^{16}$ Coulombs, it can generate a GRB. Future joint gravitational waves (GW)/GRB/FRB searches, specially after LIGO discoveries GW150914, GW151226 and LVT151012, may set some constraints on charged black holes.\\

Asymptotically anti-de Sitter (AdS) black holes have been studied extensively after Hawking-Page paper \cite{Hawking:1982dh}, in which they discussed a phase transition in the case of Schwarzschild-AdS black hole. Since then thermodynamics of more complicated AdS black holes have been investigated, a first order phase transition has been viewed in the case of  Reissner-Nordstr\"{o}m-anti de Sitter (RN-AdS) \cite{Chamblin:1999tk,Chamblin:1999hg}. Also, a second order phase transition has been realized in the case of KN-AdS \cite{Davies:1989ey}. In the mimetic $f(R)$ gravitational theory the action is adjusted by making use of a Lagrangian multiplier and mimetic potential, then a vacuum solution of RN-AdS black hole has been derived under some restrictions distinguishing this mimetic $f(R)$ variant from the $f(R)$ gravity \cite{Oikonomou:2016fxb}. It is the aim of the present study to derive a novel class of solutions of charged black holes coupled with the linear and the nonlinear electrodynamics Maxwell field in the context of  mimetic gravitational theory. We derive analytical charged black hole solutions in $D$-dimension, using the nonlinear electrodynamics. We show that the  metric contains monopole, dipole and quadruple terms. Interestingly, the dipole and quadruple terms are strongly related to some constant so that these terms vanish when the constant has a nil value, and then the solution reduces to the linear Maxwell field solution.\\

The arrangement of this study is as follow: In Section \ref{S2}, a brief review of the  mimetic  gravitational theory is given. In  Section \ref{S3}, a new black hole solution is derived. The solution behaves asymptotically as a flat spacetime.  In  Section \ref{S4}, we apply a coordinate transformation to the obtained solution, then derive analytic $D$-dimension rotating charged solutions in the framework of mimetic theory.  In Section \ref{S5}, we derive new $D$-dimension charged black hole solutions, using the nonlinear electrodynamics, which show  that the metric has monopole, dipole and quadruple terms. In Section \ref{S6}, we study the properties of the black hole solutions derived in Sections \ref{S3} and \ref{S5} by calculating the curvature invariants. This show that the black hole solutions in the nonlinear electrodynamics case have singularities stronger than those derived from  the linear case. In  Section \ref{S7}, we study the thermodynamic properties of the solutions derived in   Sections \ref{S3} and \ref{S5}.   Final section, is devoted to summarize  the present study.
\section{Preliminaries of Mimetic Gravitational Theory}\label{S2}
In this section, we discuss the case when mimetic gravity is coupled to electrodynamics in presence of a cosmological constant. The original mimetic gravity variant was first constructed in the frame of the general relativity to investigate dark matter in cosmology \cite{Chamseddine:2013kea}. The construction of the theory depends on redefining of the physical metric $g_{\mu \nu}$ \cite{Chamseddine:2013kea}:
\begin{equation} \label{par}
g_{\alpha \beta}=-({\bar g^{\mu \nu}}\partial_\mu \phi \partial_\nu \phi ){\bar g_{\alpha \beta}}, \end{equation}
where ${\bar g_{\alpha \beta}}$ is the conformal auxiliary metric,  $\phi$ is
the mimetic scalar field and  ${\bar g^{\alpha \beta}}$  is the inverse of ${\bar g_{\alpha \beta}}$. Using Eq. (\ref{par}) one can show that
GR theory is invariant under the conformal transformation, i.e.,  ${\bar g_{\alpha \beta}} \rightarrow \omega(x_\mu) {\bar g_{\alpha \beta}}$ with $\omega(x_\mu)$ being an arbitrary function of the coordinates.  Equation (\ref{par}) shows that  the mimetic field should
satisfy Eq. (\ref{con}).

Now we consider the mimetic Maxwell theory in the presence of a cosmological constant $\Lambda$. Thence, the action of this field is given by
\begin{equation} \label{act}
{\cal S}:=\frac{1}{2\chi} \int d^Dx \sqrt{-g(\bar{g}_{\mu \nu},\phi)} \left\{R(\bar{g}_{\mu \nu},\phi)
-2\Lambda\right\}-\int d^Dx \sqrt{-g(\bar{g}_{\mu \nu},\phi)}  {\cal L}_{em},
\end{equation}
where $\chi$ is the $D$-dimensional gravitational constant $\chi =2(D-3)\Omega_{D-1} G_D$ and $G_D$ is the  gravitation Newtonian  constant in $D$-dimensions, the $D$-dimensional cosmological constant $\Lambda=-\frac{(D-1)(D-2)}{2\lambda^2}$ with a length scale $\lambda$ of the dS spacetime, and ${\cal L}_{em}={ F}\wedge ^{\star}{F}$ is the Maxwell Lagrangian, with $F=dV$ and  $V=V_{\mu}dx^\mu$ being the gauge potential 1-form \cite{Awad:2017tyz, Capozziello:2012zj}. We denote the volume of $(D-1)$-dimensional unit sphere by $\Omega_{D-1}$, where
\begin{equation}
\Omega_{D-1} = \frac{2\pi^{(D-1)/2}}{\Gamma((D-1)/2)},
\end{equation}
with  $\Gamma$ being the gamma function that depends on the dimension of the space-time\footnote{For $D= 4$, one can recover
$2(D-3)\Omega_{D-1} = 8 \pi$.} and $g \equiv g(\bar{g}_{\mu \nu},\phi)$ being  the determinant of the physical metric defined by Eq. (\ref{par}).

Varying the action of Eq. (\ref{act}) with respect to the physical metric, one can derive the following equations of motion of the gravitational field
\begin{equation} \label{fe}
E_\mu{^\nu}\equiv  G_\mu{^\nu}+\frac{1}{2}\delta_{\mu }^\nu \Lambda -{\mathop{\mathcal{T}}\limits^{\textmd{\tiny em}}}_\mu{^\nu}-\widetilde{\mathcal{T}}_\mu{}^ \nu= 0,
\end{equation}
where $ G_{\mu \nu}$ is Einstein tensor, $\widetilde{\mathcal{T}}_{\mu}{^\nu}$ is the energy-momentum tensor of the mimetic field and ${\mathop{\mathcal{T}}\limits^{\textmd{\tiny em}}}{_\mu}{^\nu}$ is the energy-momentum tensor of the electromagnetic field
\begin{equation} \label{maxf1}
{\mathop{\mathcal{T}}\limits^{\textmd{\tiny em}}}{_\mu}{^\nu}=F_{\mu \alpha}F^{\nu \alpha}-\frac{1}{4} \delta_\mu^\nu F_{\alpha \beta}F^{\alpha \beta}.
\end{equation}
Notably, the auxiliary metric does not appear explicitly in  the  field equations, but it implicitly does through the physical metric given in Eq. (\ref{con}) and the mimetic field $\phi$. The presence of the mimetic field in the field equations can be written as
\begin{equation} \label{ten1}
\widetilde{\mathcal{T}}_{\mu \nu}=-\left(G+\frac{D}{2}\Lambda-{\mathop{\mathcal{T}}\limits^{\textmd{\tiny em}}}\right)\partial_\mu\phi\, \partial_\nu\phi, \end{equation}
where $G = -\left(\frac{D-2}{2}\right)R$ is the trace of Einstein tensor. Finally, the variation in term of the action (\ref{act}) with respect to the vector potential $V_{\mu}$ yields \cite{Awad:2017tyz}
\begin{equation} \label{maxf3}
\partial_\nu \left( \sqrt{-g} F^{\mu \nu} \right)=0.\end{equation}

It is worth to mention that the energy-momentum tensors, ${\mathop{\mathcal{T}}\limits^{\textmd{\tiny em}}}_{\mu \nu}$ and $\widetilde{\mathcal{T}}_{\mu \nu}$ , are  conserved, i.e. satisfy the continuity equations $\nabla^\mu {\mathop{\mathcal{T}}\limits^{\textmd{\tiny em}}}_{\mu \nu}=0=\nabla^\mu \widetilde{\mathcal{T}}_{\mu \nu}$, where $\nabla$ is the covariant derivative. Using the mimetic field constraint (\ref{con}) and the energy-momentum tensor (\ref{ten1}), the corresponding continuity reads
\begin{equation} \label{cons3}
\nabla^\kappa\left(\left[G+\frac{D}{2}\Lambda-{\mathop{\mathcal{T}}\limits^{\textmd{\tiny em}}}\right]\partial_\kappa \phi\right)= \frac{1}{\sqrt{-g}}\partial_\kappa\left(\sqrt{-g}\left[G+\frac{D}{2}\Lambda-{\mathop{\mathcal{T}}\limits^{\textmd{\tiny em}}}\right] g^{\kappa \sigma}\partial_\sigma \phi \right)=0.
\end{equation}
Alternatively, one finds that (\ref{con}) is satisfied identically, when (\ref{cons3}) is used. It is straightforward to show that the trace of Eq. (\ref{fe}) has the form
\begin{equation}
\left[G+\frac{D}{2}\Lambda-{\mathop{\mathcal{T}}\limits^{\textmd{\tiny em}}}\right]\left(1+g^{\mu \nu}\partial_\mu\phi \partial_\nu \phi\right)\equiv 0,
\end{equation}
which is satisfied identically due to the mimetic field constraint, namely Eq. (\ref{con}). In conclusion, we note that the conformal degree of freedom provides a dynamical quantity, i.e.  ($G \neq  0$), and therefore the mimetic theory has non-trivial solutions for the conformal mode even in the absence of matter \cite{Chamseddine:2013kea}.

\section{$D$-dimension Charged Black Holes in Mimetic Gravity }\label{S3}
In this section, we present $D$-dimension solution of a charged black hole in mimetic gravity. So we suppose the spacetime configuration is given by the metric
\begin{equation} \label{m2}
ds^2= -f(r)dt^2+\frac{1}{f(r)}dr^2+r^2\left(\sum_{i=1}^{\ell}d\theta^2_i+\sum_{k=1}^{D-\ell-2}dz_k^2\right),\end{equation}
where $0\leq r< \infty$, $-\infty < t < \infty$, $0\leq \theta_{\ell}< 2\pi$ and $-\infty < z_k < \infty$. Here $f(r)$ is unknown function  of the radial coordinate $r$ only. For the spacetime (\ref{m2}), we get the Ricci scalar
\begin{equation}\label{ri}
R=-\frac{r^2 f''+2(D-2)r f'+(D-2)(D-3)f}{r^2}.
\end{equation}
Then, the non-vanishing components of Eqs. (\ref{fe}) and (\ref{maxf1}) read the following set of field equations
\begin{eqnarray} \label{fe13}
& & E_t{}^t=\frac{(D-2)rf'+(D-2)(D-3)f+\Lambda r^2+q'^2r^2+fn_{z_{(D-3)}}^2-2fn_{z_{(D-3)}}s'+fs'^2}{2r^2}\equiv 0,\nonumber\\
& & E_t{}^{z_{(D-3)}}=E_{{z_{(D-3)}}}{}^t= q'[n_{z_{(D-3)}}-s']\equiv 0,\nonumber\\
& & E_r{}^r= \frac{1}{2r^2}\Bigg\{(D-2)rf'+(D-2)(D-3)f+\Lambda r^2-\phi'^2f[2(D-2)^2f' r+(D-2)^2(D-3)f+(D-2)f''r^2+6\Lambda r^2]-q'^2r^2\nonumber\\
& & -f n_{z_{(D-3)}}^2+2fn_{z_{(D-3)}}s'-fs'^2+(D-4)f\phi'^2[q'^2r^2+2fs'n_{z_{(D-3)}}-fs'^2-fn_{z_{(D-3)}}^2]\Bigg\}\equiv 0,\nonumber\\
& & E_{\theta_1}{}^{\theta_1} =E_{\theta_2}{}^{\theta_2}= \cdots =E_{_{\theta_{\ell}}}{}^{_{ \theta_{\ell}}}= \frac{1}{2r^2}\left[2(D-3)rf'+(D-4)(D-3)f+r^2f''+\Lambda r^2-r^2q'^2-2fs'n_{z_{(D-3)}}+fs'^2+fn_{z_{(D-3)}}^2\right]\equiv 0,\nonumber\\
& & E_{z_1}{}^{z_1}=E_{z_2}{}^{z_2}=\cdots = E_{z_{D-\ell-2}}{}^{_{z_{D-\ell-2}}}=\frac{1}{2r^2}\left[2(D-3)rf'+(D-4)(D-3)f+r^2f''+\Lambda r^2-r^2q'^2+ 2fs'n_{z_{(D-3)}}-fs'^2-fn_{z_{(D-3)}}^2\right]\equiv 0,\nonumber\\
& &
\end{eqnarray}
where $f'=\frac{df(r)}{dr}$, $s'=\frac{ds(r)}{dr}$,  $q'=\frac{dq(r)}{dr}$ and $n_{z_{(D-3)}}=\frac{dn(z_{(D-3)})}{dz_{(D-3)}}$ with $q(r)$, $n(z_{(D-3)})$  and $s(r)$ being three unknown functions related to the electric and magnetic charges of the black hole. These are usually defined from the general form of the vector potential
\begin{equation}\label{p}
V =q(r)dt+n(z_{(D-3)})dr+s(r)dz_{(D-3)}.
\end{equation}
Notably, for vanishing $n(z_{(D-3)})$ and $s(r)$, one can generate $D$-dimension charged electric solutions in the mimetic gravitational theories. However, for non-vanishing $q(r)$, $n(z_{(D-3)})$ and $s(r)$, one expects rich physical properties to showup. We solve the field equations (\ref{fe13}) as follows
\begin{eqnarray} \label{so1}
& & q(r)=\frac{c_1}{r^{D-3}}\; ,\qquad n(z_{(D-3)})=c_2z_{(D-3)}\;,\qquad s(r)=c_3r\; , \qquad f(r)=-\frac{\Lambda r^2}{(D-1)(D-2)}+\frac{c_4}{r^{D-3}}+\frac{(D-3)c_1{}^2}{(D-2)r^{2(D-3)}} \;,\nonumber\\
&&  \phi(r)=\pm\left[f_1(r)+c_4 \ln{\sqrt{r-c_4/2+f_1(r)}}\right]\delta^4_4+c_5, \qquad \textrm{with} \qquad f(r)=\frac{1}{f_1(r)}\left(2f_1(r)-(r-c_4)\tan^{-1}\left[\frac{f_2(r)}{2{f_1(r)}}\right]\right),\nonumber\\
&& f_1(r)=\sqrt{r^2-c_4r}, \qquad f_2=2r-c_4.
 \end{eqnarray}
where $c_i$, $i=1\cdots 5$ are constants. It is worth to mention that Eq. (\ref{so1}) is an exact solution of Maxwell-mimetic gravitational theory given by Eqs. (\ref{fe}) and (\ref{maxf1}) in addition to the trace which given by  Eq. (\ref{ten1}). As clear from the obtained solution (\ref{so1}) that the mimetic field $\phi(r)$ is a function of the radial coordinate $r$ in the $4$-dimensional spacetime case, while in the case $D>4$ the mimetic field becomes constant. Plugging the solution (\ref{so1}) into the spacetime metric (\ref{m2}), we write
\begin{eqnarray} \label{m51}
&&ds^2=-\left\{-\frac{\Lambda r^2}{(D-1)(D-2)}+\frac{c_4}{r^{D-3}}+\frac{(D-3)c_1{}^2}{(D-2)r^{2(D-3)}}\right\}dt^2+\left\{-\frac{\Lambda r^{D-1}}{(D-1)(D-2)}+\frac{c_4}{r^{D-3}}+\frac{(D-3)c_1{}^2}{(D-2)r^{2(D-3)}}\right\}^{-1}dr^2\nonumber\\
&& +r^2\left(\sum_{i=2}^{\ell}d\theta^2_i
+\sum_{k=2}^{D-\ell-2}dz_k^2\right).
\end{eqnarray}
As clear the spacetime (\ref{m51}) is asymptotically (A)dS, also it is obvious that the magnetic fields related to $n(z_{(D-3)})$ and  $s(r)$ do not contribute to the metric.
\section{Rotating Black String Solutions}\label{S4}
In this section, we derive rotating solutions satisfying the field equations (\ref{fe}) and (\ref{maxf1}) of Maxwell-mimetic theory. In order to do so, we apply the coordinate transformations
\begin{equation} \label{t1}
{\theta}'_i =-\Xi~ {\theta}_i+\frac{ \omega_i}{\lambda^2}~t,\qquad \qquad \qquad {t}'= \Xi~ t-\sum\limits_{i=1}^{{\ell}} \omega_i~ \theta_i, \end{equation} where $\omega_i$, $i\geq1$ is the number of rotation parameters and $\Xi$ is defined as
\[\Xi:=\sqrt{1+\sum\limits_{i=1}^{{\ell}}\frac{ \omega_i}{\lambda^2}}.\]
The rotation group in $D$-dimensions is so ($D-1$) and the independent number of the rotation parameters
for a localized object is equal to the number of Casimir operators, which is [$(D-1)/2$], where $[y]$ is the
integer part of $y$. Applying the transformation (\ref{t1}) to the metric (\ref{m51}),  we get
\begin{eqnarray} \label{rot}
& & ds^2=-f(r)\left[\Xi d{t'}  -\sum\limits_{i=1}^{\ell}  \omega_{i}d{\theta'}_i \right]^2+\frac{dr^2}{f(r)}+\frac{r^2}{\lambda^4}\sum\limits_{i=1 }^{\ell}\left[\omega_{i}d{t'}-\Xi \lambda^2 d{\theta'}_i\right]^2+ r^2 \sum\limits_{k=1}^{D-\ell-2}dz_k^2-\frac{r^2}{\lambda^2}\sum\limits_{i<j }^{\ell}\left(\omega_{i}d{\theta'}_j-\omega_{j}d{\theta'}_i\right)^2,\end{eqnarray}  where $f(r)$ is given by Eq. (\ref{so1}). We note that the static configuration (\ref{m51}) can be recovered as a special case when the rotation parameters $\omega_\ell$ are made to vanish. Also, it is important to mention that the vanishing of the quantities $c_1$ and $c_4$ leads to an odd  AdS spacetime. On the other hand, it is easy task to show that the limiting metric is a Minkowski spacetime, since all curvature components vanish identically.

In general the coordinate transformation (\ref{t1}) is admitted only locally \cite{Lemos:1994xp,Awad:2002cz}, since it relates time to periodic coordinate $\theta_\ell$. On other words, the spacetimes (\ref{rot}) and (\ref{m2}) can be locally mapped into each other but not globally and thus they are different. This has been discussed in more detail in \cite{Stachel:1981fg}, for similar coordinate transformation, it has been shown that if the first Betti number of a manifold has a non-vanishing value, then there are no global diffeomorphisms can connect the two spacetimes. Therefore, the manifold parameterized globally by the rotation parameters $\omega_\ell$ is different from the static spacetime. The solution (\ref{so1}) shows that the first Betti number is one, which characterizes the cylindrical or toroidal horizons.
\section{New Black Holes with Nonlinear Electrodynamics in Mimetic Gravity }\label{S5}
In this section, we consider the mimetic theory with nonlinear electrodynamics in the presence of a cosmological constant.  Therefore, we take the action
\begin{equation} \label{act1}
{\cal S_{NL}}:=\frac{1}{2\chi} \int d^Dx \sqrt{-g(\bar{g}_{\mu \nu},\phi)} \left\{R(\bar{g}_{\mu \nu},\phi)-2\Lambda\right\}-\int  d^{D}x\sqrt{-g(\bar{g}_{\mu \nu},\phi)} {\cal L(F)},\end{equation}
where ${\cal L(F)}$ is the Lagrangian of the nonlinear electrodynamics. Alternatively, we could reexpress ${\cal L(F)}$ in terms of Legendre transformation
\begin{equation}
{\cal Q}=2F {\cal L}_F-{\cal L}, \quad \textrm{where}\qquad {\cal L}_F=\frac{\partial {\cal L}}{\partial F}.
\end{equation}
It is useful to define the second-order tensor
\begin{equation} \label{pn} P_{\mu \nu}={\cal L}_F F_{\mu \nu},\end{equation}
where the linear Maxwell field is recovered by setting ${\cal L}_F=1$. As clear from the above that ${\cal Q}$ is a function of $P$, where \cite{Salazar:1987ap, AyonBeato:1999rg}
\begin{equation} P=\frac{1}{4}P_{\mu \nu}P^{\mu \nu}=[{\cal L}_F]^2F.\end{equation}
Thus, we have
$$d{\cal Q}=\frac{1}{{\cal L}_F}d\left\{[{\cal L}_F]^2F \right\}={\cal Q}_PdP, \qquad {\textmd {where}} \qquad {\cal Q}_P=\frac{\partial {\cal Q}}{\partial P}.$$

Varying the action of Eq. (\ref{act1}) with respect to the physical metric, one can write the gravitational field equations
\begin{equation}\label{fe1}
E_{\mu }{^\nu} = G_{\mu }{^\nu}+\frac{1}{2}\delta_\mu ^\nu \Lambda-
\mathop{\mathfrak{T}}\limits^{\textmd{\tiny NL}}{_\mu}{^\nu}-\widetilde{\mathfrak{T}}{_{\mu}}{^{\nu}}\equiv0,
\end{equation}
and Maxwell field equations of the nonlinear electrodynamics \cite{AyonBeato:1999rg}
\begin{equation} \label{maxf}
\partial_\nu \left( \sqrt{-g} P^{\mu \nu} \right)=0.\end{equation}
In the above, the energy-momentum tensor,
\begin{equation} \label{max1}
{\mathop{\mathfrak{T}}\limits^{\textmd{\tiny NL}}}{_\mu}{^\nu}=2({\cal Q}_PP_{\mu \alpha}P^{\nu \alpha}-\delta_\mu^\nu [2P{\cal Q}_P-{\cal Q}]), \end{equation}
whose a non-vanishing trace. In addition, the mimetic field contributes in the field equations as
\begin{equation} \label{ten}
\widetilde{\mathfrak{T}}_{\mu \nu}=-\left(G+\frac{D}{2}\Lambda-{\mathop{\mathfrak{T}}\limits^{\textmd{\tiny NL}}}\right)\partial_\mu\phi\, \partial_\nu\phi.
\end{equation}

Similar to the linear case, the energy-momentum tensors, ${\mathop{\mathfrak{T}}\limits^{\textmd{\tiny NL}}}_{\mu \nu}$ and $\widetilde{\mathfrak{T}}_{\mu \nu}$ are conserved, i.e. $\nabla^\mu {\mathop{\mathfrak{T}}\limits^{\textmd{\tiny NL}}}_{\mu \nu}=0=\nabla^\mu \widetilde{\mathfrak{T}}_{\mu \nu}$.  Using Eqs. (\ref{con}) and (\ref{ten}), we write the continuity equation of the mimetic field
\begin{equation} \label{cons}
\nabla^\kappa\left(\left[G+\frac{D}{2}\Lambda-{\mathop{\mathfrak{T}}\limits^{\textmd{\tiny NL}}}\right]\partial_\kappa \phi\right)\equiv \frac{1}{\sqrt{-g}}\partial_\kappa\left(\sqrt{-g}\left[G+\frac{D}{2}\Lambda-{\mathop{\mathfrak{T}}\limits^{\textmd{\tiny NL}}}\right] g^{\kappa \sigma}\partial_\sigma \phi \right)=0.
\end{equation}
It is useful to give the trace of Eq. (\ref{fe1})
\begin{equation} \label{cons1}
\left(G+\frac{D}{2}\Lambda-{\mathop{\mathfrak{T}}\limits^{\textmd{\tiny NL}}}\right)(1+g^{\mu \nu}\partial_\mu\phi \partial_\nu \phi)=0.
\end{equation}
As clear, Eq. (\ref{cons1}) is satisfied identically, if  Eq. (\ref{con}) is used. Since $G \neq  0$ or ${\mathop{\mathfrak{T}}\limits^{\textmd{\tiny NL}}} \neq  0$, the mimetic theory at hand has non-trivial solutions and the conformal degree of freedom, remarkably, provides a dynamical quantity \cite{Chamseddine:2013kea}.

We apply the field equations (\ref{fe1}) and (\ref{maxf}) to the spacetime metric (\ref{m2}), where the non-vanishing components are
\begin{eqnarray} \label{fe3}
& & E_t{}^t\equiv\frac{(D-2)rf'+(D-2)(D-3)f+\Lambda r^2-2r^2{\cal Q(r)}}{2r^2}=0,\nonumber\\
& & E_r{}^r\equiv \frac{1}{2fr^2}\Bigg\{(D-2)rf'+(D-2)(D-3)f+\Lambda r^2-\phi'^2f[2(D-2)^2f' r+(D-2)^2(D-3)f+(D-2)f''r^2+D\Lambda r^2-2Dr^2{\cal Q(r)}]\nonumber\\
& &-2r^2{\cal Q(r)}-2(D-2)\phi'^2f\varphi'\varphi''^{-1}r^2{\cal Q(r)}'\Bigg\}=0,\nonumber\\
& &E_{\theta_1}{}^{\theta_1} =E_{\theta_2}{}^{\theta_2}= \cdots =E_{_{\theta_{\ell}}}{}^{_{ \theta_{\ell}}}= E_{z_1}{}^{z_1}=E_{z_2}{}^{z_2}=\cdots = E_{z_{D-\ell-2}}{}^{_{z_{D-\ell-2}}}\nonumber\\
& & \equiv \frac{1}{2r^2\varphi''}\left[2(D-3)rf'\varphi''+\varphi''\{(D-4)(D-3)f+r^2f''+\Lambda r^2-2r^2{\cal Q(r)}\}+2r^2\varphi'Q'\right]=0.
\end{eqnarray}
In the above, ${\cal Q(r)}$ is an arbitrary function representing nonlinear electrodynamics, and $\varphi(r)$ is an unknown function reproducing $D$-dimension electric charge in the vector potential where $P=d\cal{V}$ and  $\cal {V}=\cal{V}_{\mu}dx^\mu$ being the gauge potential 1-form  which defined  in the non-linear case as
\begin{equation} \cal{ V}=\varphi(r)dr.\end{equation}
 The general solution  of the system of differential equations (\ref{fe3}) have the form
\begin{eqnarray} \label{so5}
& &{\cal Q(r)}=\frac{c_6}{r^{2(D-2)}}+\frac{c_7{}}{r^{(2D-3)}}+\frac{c_7^2}{r^{3(D-2)}}\; ,\quad  \phi(r)=\left[f_1(r)+c_9 \ln{\sqrt{r-c_9/2+f_1(r)}}\right], \qquad \textrm{with}  \quad f_1(r)=\sqrt{r^2-c_9r},\nonumber\\
&&   \varphi(r)=\frac{c_8}{r^{D-3}}, \qquad f(r)=\frac{\Lambda r^2}{(D-1)(D-2)}+\frac{c_9}{r^{D-3}}-\frac{2c_6}{(D-2)(D-3)r^{2(D-3)}}-\frac{2c_7}{(D-2)^2r^{2D-5}}-\frac{4c_7^2}{(D-2)(2D-5)r^{(3D-8)}} \;,\nonumber\\[15pt]
 \end{eqnarray}
where $c_i$, $i=6\cdots 9$ are constants. From the above solutions, we can conclude: In the solution (\ref{so5}) as clear from the $\mathcal{Q}(r)$ function that the $4$-dimension nonlinear electrodynamics consists of monopole, dipole and quadrupole terms. Interestingly, the constant associated with the monopole is different from that appears with the dipole and quadrupole terms. In this case, one can get the form of the nonlinear electrodynamics by setting the constant $c_7=0$. On the other hand, the mimetic field $\phi(r)$ is constant in this case.

It is worth to mention that Eq. (\ref{so5}) is an exact solution to Maxwell-mimetic gravitational theory that is given by Eqs. (\ref{fe1}) and (\ref{maxf}). Then, we write explicitly the line elements corresponding to the above solution   as

\begin{eqnarray}
ds^2&=&-\left\{\frac{\Lambda r^2}{(D-1)(D-2)}+\frac{c_9}{r^{D-3}}-\frac{2c_6}{(D-2)(D-3)r^{2(D-3)}}-\frac{2c_7}{(D-2)^2r^{2D-5}}-\frac{4c_7{}^2}{(D-2)(2D-5)r^{(3D-8)}}\right\}dt^2\nonumber\\
 &&+\left\{\frac{\Lambda r^2}{(D-1)(D-2)}+\frac{c_9}{r^{D-3}}-\frac{2c_6}{(D-2)(D-3)r^{2(D-3)}}-\frac{2c_7}{(D-2)^2r^{2D-5}}-\frac{4c_7{}^2}{(D-2)(2D-5)r^{(3D-8)}}\right\}^{-1}dr^2\nonumber\\
& &+r^2\left(\sum_{i=2}^{\ell}d\theta^2_i+\sum_{k=2}^{D-\ell-2}dz_k^2\right)\;,\label{m5}
\end{eqnarray}
One can easily recognize that the spacetime (\ref{m5}) is asymptotically (A)dS.

Similar to what has been done in Sec. \ref{S4}, we add angular momentum to the spacetime (\ref{m5}) by applying the transformation (\ref{t1}). Therefore, the metric becomes
\begin{eqnarray} \label{m1}
& & ds^2=-f(r)\left[\Xi d{t'}  -\sum\limits_{i=2}^{\ell}  \omega_{i}d{\theta'} \right]^2+\frac{dr^2}{f(r)}+\frac{r^2}{\lambda^4}\sum\limits_{i=1 }^{\ell}\left[\omega_{i}d{t'}-\Xi \lambda^2 d{\theta'}_i\right]^2+ r^2 \sum\limits_{k=1}^{D-\ell-2}dz_k^2-\frac{r^2}{\lambda^2}\sum\limits_{i<j }^{\ell}\left(a_{i}d{\phi'}_j-a_{j}d{\phi'}_i\right)^2,
\end{eqnarray}
where $f(r)$ is given by Eq. (\ref{so5}). We note that coupling gravity and nonlinear electrodynamics has been studied in gravitational theories other than mimetic gravity, c.f \cite{Berej:2006cc,Myung:2007av}. In \cite{Myung:2007av}, GR gravity is coupled to nonlinear electrodynamics to study thermodynamics of magnetically charged black holes. Also, in \cite{Berej:2006cc}, quadratic curvature gravity is coupled with nonlinear electrodynamics to obtain regular black hole solutions, in that work the perturbative solution has been derived up to first order, where the regularity of the obtained solution depends on two free parameters of the model. In both models, nonlinear electrodynamic contribution has been assumed to have a particular form to reduce to RN asymptotically, then the solutions have been adjusted to satisfy the field equations. This is in contrast to the model at hand, where the solutions, namely (\ref{so5}), have been obtained directly from the equations of motion without pre-assumptions of the nonlinear electrodynamics contribution.
\section{Features of the Black Hole Solutions }\label{S6}
Now we are going  to discuss  some relevant features of the charged black hole solutions presented in previous sections for the linear and nonlinear cases.

Firstly, in the linear Maxwell field case, the metric of the charged black hole solution (\ref{m51}) takes the form
\begin{eqnarray} \label{m11}
ds^2=-\Bigg[r^2\Lambda_{eff} -\frac{m}{r^{D-3}}+\frac{(D-3)q^2}{(D-2)r^{2(D-3)}}\Bigg]dt^2+\frac{dr^2}{\Biggr[r^2\Lambda_{eff} -\frac{m}{r^{D-3}}+\frac{(D-3)q^2}{(D-2)r^{2(D-3)}}\Biggr]^{-1}} +r^2\left(\sum_{i=2}^{\ell}d\theta^2_i+\sum_{k=2}^{D-\ell-2}dz_k^2\right),
\end{eqnarray}
where $\Lambda_{eff}=-\frac{\Lambda }{(D-1)(D-2)}$, $c_1=q$ and $c_4=m$. The above equation shows clearly that the metric in the linear case is the Reissner-Nordstr\"om solution which behaves asymptotically as dS background. By taking the limit $q \rightarrow 0$, we get the dS non-charged black holes. However, the horizons of the metric (\ref{m11}) are given by the real positive roots of $\Gamma(r)=0$, where $\Gamma(r)={r^{4(D-3)}}\Lambda -mr^{(D-3)}+\frac{(D-3)q^2}{(D-2)}$, see \cite{Brecher:2004gn}. For the model at hand, namely the spacetime metric (\ref{m11}), taking $x=r^2$, we find that the constraint $\Gamma(x)=0$ gives two positive real roots in the $4$-dimension case, one of the roots represents the black hole event (inner) horizon, $r_b$ and the other one represents a cosmological (outer) horizon, $r_c$. Similarly, in the $5$-dimension case, the constraint  $\Gamma(x)=0$ gives three positive roots, i.e. the solution has three horizons. In general, in the $D$-dimension case, the constraint $\Gamma(x)=0$ gives $(D-2)$ horizons.

Secondly, in the the nonlinear Maxwell field case, the spacetime metrics of the charged black hole solutions (\ref{m5})  takes the form
\begin{eqnarray}
ds^2&=&-\left\{\Lambda_{eff} r^2-\frac{m}{r^{D-3}}+\frac{q^2}{(D-2)(D-3)r^{2(D-3)}}+\frac{q_1}{(D-2)^2r^{2D-5}}+\frac{q_1{}^2}{4(D-2)(2D-5)r^{(3D-8)}}\right\}dt^2\nonumber\\
 &&+\left\{\Lambda_{eff} r^2-\frac{m}{r^{D-3}}-\frac{q^2}{(D-2)(D-3)r^{2(D-3)}}+\frac{q_1}{(D-2)^2r^{2D-5}}+\frac{q_1{}^2}{4(D-2)(2D-5)r^{(3D-8)}}\right\}^{-1}dr^2\nonumber\\
 &&+r^2\left(\sum_{i=2}^{\ell}d\theta^2_i+\sum_{k=2}^{D-\ell-2}dz_k^2\right)\;,\label{m22}
\end{eqnarray}
where $c_9=-m$, $c_6=-q^2/2$ and  $c_7=-q_1/2$.  Equation  (\ref{m22}) shows clearly that the metric of the charged black hole in the nonlinear Maxwell case is different from RN black hole. This difference is due to the existence of the dipole and quadrupole terms that are related by the constant $q_1$. However, in the case that $q_1=0$, the solution reduces to the RN case. It is of interest to note that all the charged terms appear in Eq. (\ref{m22}) are reproduced from the arbitrary function, $\mathcal{Q}(r)$, which characterizes the nonlinear electrodynamics. By taking the limit $q_1\rightarrow 0$, the solution goes to the (A)dS charged black hole of the linear case.  Additionally, we investigate the number of horizons of the spacetime metric (\ref{m22}). In the $4$-dimension case, by taking $x=r^3$, similar to the linear electrodynamics case, we find that the constraint $\Gamma(x)=0$ has two positive real roots, one for the black hole even horizon, $r_b$, and the other is for a cosmological horizon, $r_c$. In the $5$-dimension case, the constraint $\Gamma(x)=0$ gives three positive roots and in the general $D$-dimension case, the constraint $\Gamma(x)=0$ gives $(D-2)$ horizons.

Next we discuss some physical properties of the solutions (\ref{so1}) and (\ref{so5}) by investigating the singularity behaviours and the their stabilities.
\subsection{Visualization of black holes' singularities}
We investigate the physical singularities by calculating some the curvature invariants. In both cases, linear and nonlinear electrodynamics, see (\ref{so1}) and (\ref{so5}), it is clear that the function $f(r)$ is irregular when $f_1(r)=0$. Therefore, it is worth to investigate the regularity of the solutions at $f_1(r)=0$. For the solutions (\ref{so1}), we evaluate the scalar invariants
\begin{equation} \label{scal1}
R^{\mu \nu \lambda \rho}R_{\mu \nu \lambda \rho}= \frac{L_1(r)}{r^{4(D-2)}}, \qquad \qquad R^{\mu \nu}R_{\mu \nu}=\frac{L_2(r)}{r^{4(D-2)}}, \qquad \qquad R= \frac{L_3(r)}{r^{2(D-2)}},
\end{equation}
where $R^{\mu \nu \lambda \rho}R_{\mu \nu \lambda \rho}$, $R^{\mu \nu}R_{\mu \nu }$, $R$ are the Kretschmann scalars,  the Ricci tensors square, the Ricci scalars, respectively, also $L_i(r)$ are some polynomial functions in the radial coordinate $r$. Equation (\ref{scal1}) shows that the solutions, at $r=0$, have true singularities.

For the solutions (\ref{so5}), we evaluate the invariants
\begin{equation} \label{scal}
R^{\mu \nu \lambda \rho}R_{\mu \nu \lambda \rho}= \left(\frac{L_4(r)}{r^{6(D-2)}}\right), \qquad R^{\mu \nu}R_{\mu \nu}=\left(\frac{L_5(r)}{r^{6(D-2)}}\right), \qquad  R= \left(\frac{L_6(r)}{r^{3(D-2)}}\right).
\end{equation}
As clear all invariants have true singularities at $r=0$. Remarkably, at the limit $r\to 0$, the behaviours of the Kretschmann scalars, the Ricci tensor square and the Ricci scalars of the nonlinear charged black hole solutions are given by $K=R_{\mu \nu}R^{\mu \nu} \sim r^{-6(D-2)}$, $R\sim  r^{-3(D-2)}$ in contrast with the  solutions of the linear Maxwell mimetic theory which have $K =R_{\mu \nu}R^{\mu \nu}\sim r^{-4(D-2)}$ and $R\sim  r^{-2(D-2)}$. This shows clearly that the singularities in the nonlinear electrodynamics case is stronger than that are has been obtained in the linear Maxwell mimetic gravity case. Notably, one may check if geodesics are extendible beyond these regions. According to Tipler and Kr\'{o}lak \cite{Tipler:1977,CLARKE1985127}, this indicates the strength of the singularity. This topic will be discussed in forthcoming studies.
\subsection{Energy conditions}
Energy conditions provide important tools to examine and better understand cosmological models and/or strong gravitational fields. We are interested in the study of energy conditions in the nonlinear electrodynamics case, since the linear case is well known in the general relativity theory. As mentioned before, we focus on the nonlinear electrodynamics black hole solutions (\ref{so5}). The Energy conditions are classified into four categories: The strong energy (SEC),  the weak energy (WEC), the null energy (NEC) and the dominant energy conditions \cite{Hs73, Nashed:2016gwp}. To fulfill those, the inequalities below must be satisfied
 \begin{eqnarray} \label{ec}
& &SEC : \rho+p_r\geq 0, \quad \rho+p_r+2p_t\geq 0, \nonumber\\
 &&WEC : \rho\geq 0, \quad \rho+p_r\geq 0,\quad \rho+p_t\geq 0, \nonumber\\
 &&NEC : \rho+p_r\geq 0, \quad \rho+p_t\geq 0, \nonumber\\
 && DEC : \rho\geq |p_r|, \quad \; \; \; \rho\geq |p_t|,
\end{eqnarray}
where ${\mathop{\mathfrak{T}}\limits^{\textmd{\tiny NL}}}{_0}{^0}=\rho$, ${\mathop{\mathfrak{T}}\limits^{\textmd{\tiny NL}}}{_1}{^1}=p_r$ and ${\mathop{\mathfrak{T}}\limits^{\textmd{\tiny NL}}}{_2}{^2}={\mathop{\mathfrak{T}}\limits^{\textmd{\tiny NL}}}{_3}{^3}=p_t$ are the density, radial and tangential pressures, respectively.
Straightforward calculations of black hole solutions (\ref{so5}), in $4$-dimensions, give
\begin{eqnarray} \label{cal}
& &SEC : \rho+p_r= \frac{4q^2r^2-4q_1r+q_1^2}{8r^6}> 0, \quad \rho+p_r+2p_t=-\frac{q_1(2r+q_1)}{8r^6}< 0, \nonumber\\
 &&WEC : \rho=\frac{4q^2r^2-4q_1r+q_1^2}{16r^6}> 0, \quad \rho+p_r=\frac{4q^2r^2-4q_1r+q_1^2}{8r^6}> 0,\quad \rho+p_t=\frac{8q^2r^2+2q_1r-q_1^2}{16r^6}> 0, \nonumber\\
 &&NEC : \rho+p_r=\frac{4q^2r^2-4q_1r+q_1^2}{8r^6}>  0, \quad \rho+p_t=\frac{8q^2r^2+2q_1r-q_1^2}{8r^6}> 0, \nonumber\\[10pt]
 && DEC : \rho\geq |p_r| \quad \textrm{(satisfied)}, \quad \; \; \; \rho\geq |p_t| \quad \textrm{(satisfied)}.
\end{eqnarray}
This shows that the SEC is the only condition that is not satisfied. Remarkably, the SEC breaking is due to the contribution of the charge  $q_1$ which characterizes the nonlinear function ${\cal Q(r)}$. This shows clearly how the nonlinear contribution of the charge $q_1$ strengthens the singularity as discussed in the previous subsection. The SEC has a relatively clear geometrical  interpretation which is  the convergence of timelike geodesics which is manifestly related to the construction  of singularities. We remind that the mimetic theory is stable if the density of the mimetic field is positive, which favors the solutions with de Sitter backgrounds in presence of a positive cosmological constant as in the work at hand \cite{Barvinsky:2013mea} (see also \cite{Chaichian:2014qba}). In conclusion, the validity of the WEC in-return guarantees the positivity of the energy density in agreement with stability condition we just mentioned.
\section{Thermodynamical Stability and Phase Transition}\label{S7}
Black hole thermodynamics is one of the most exciting topics in physics, since it subjugates the laws of thermodynamics to understand gravitational/quantum  physics of black holes. Two main approaches have been proposed to extract thermodynamical quantities of black holes: The first approach was given by Gibbons and Hawking \cite{Hunter:1998qe,Hawking:1998ct} to study thermal properties of the Schwarzschild black hole solution using the Euclidean continuation. The second approach is to identify gravitational surface and then  define temperature then study the stability of the black hole \cite{Bekenstein:1972tm,Bekenstein:1973ur,Gibbons:1977mu}.  Here in this study, we follow the second approach to study the thermodynamics of the (A)dS black holes derived in Eq. (\ref{m11}) and (\ref{m22})  then discussed their stability. These black holes are portrayed by the mass, $m$, the charges (monopole, $q$, dipole and quadrupole, $q_1$) and also by the cosmological constant $\Lambda$.

\begin{figure*}
\centering
\subfigure[~The linear electrodynamics case]{\label{fig:1a}\includegraphics[scale=0.35]{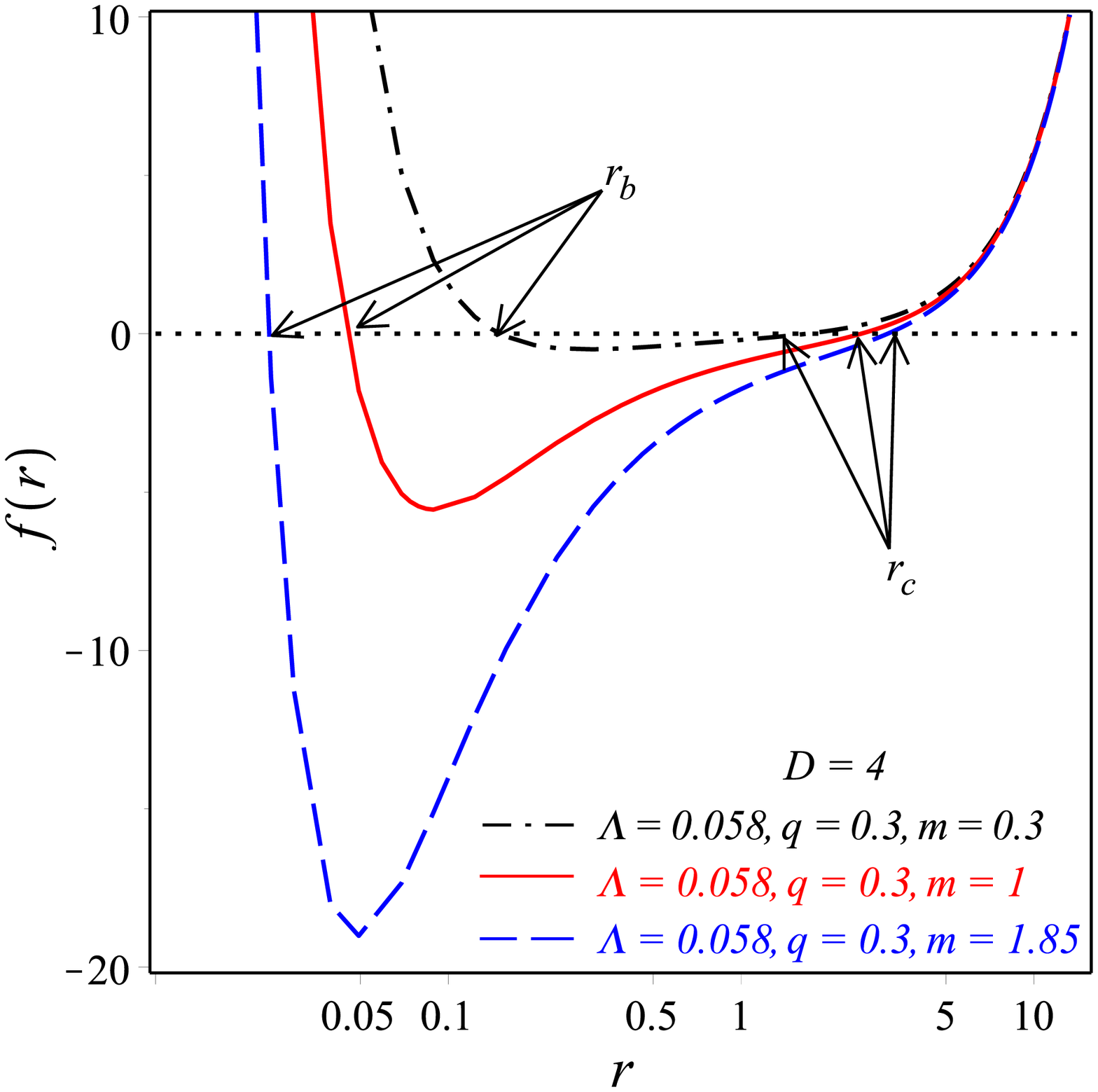}}\hspace{0.5cm}
\subfigure[~The nonlinear electrodynamics case]{\label{fig:1b}\includegraphics[scale=0.35]{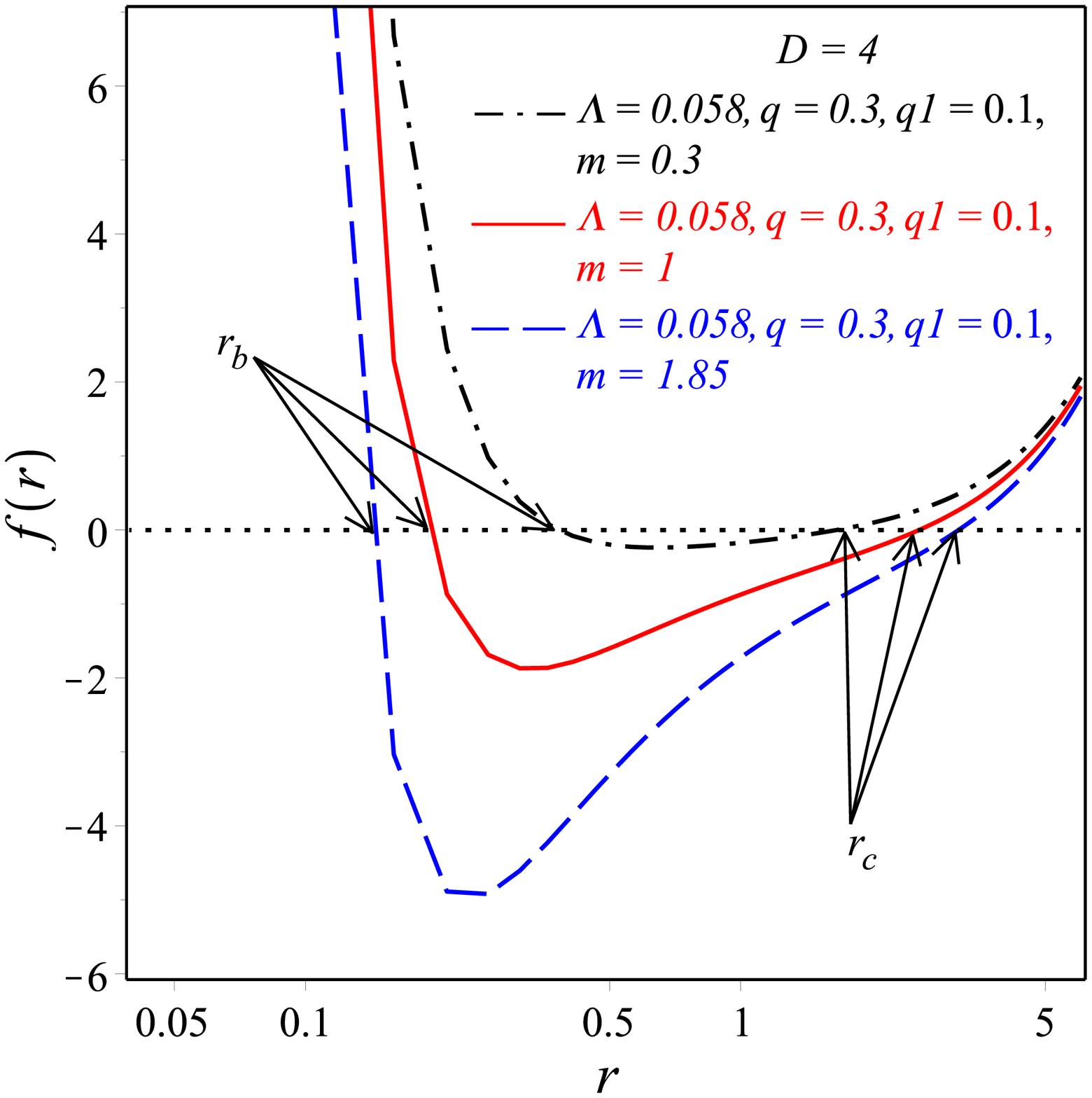}}
\caption{Schematic plots of the function $f(r)$ which characterizes the event horizons by setting $f(r)=0$: \subref{fig:1a} For the linear Maxwell mimetic gravity theory, the function $f(r)$ is given by (\ref{so1}); \subref{fig:1b} For the nonlinear electrodynamics mimetic gravity theory, the function $f(r)$ is given by (\ref{so5}). Also the plots indicate stronger singularities, as $r\to 0$, in the nonlinear electrodynamics case.}
\label{Fig:1}
\end{figure*}

\begin{figure*}
\centering
\subfigure[~possible horizons of the solution]{\label{fig:2a}\includegraphics[scale=0.35]{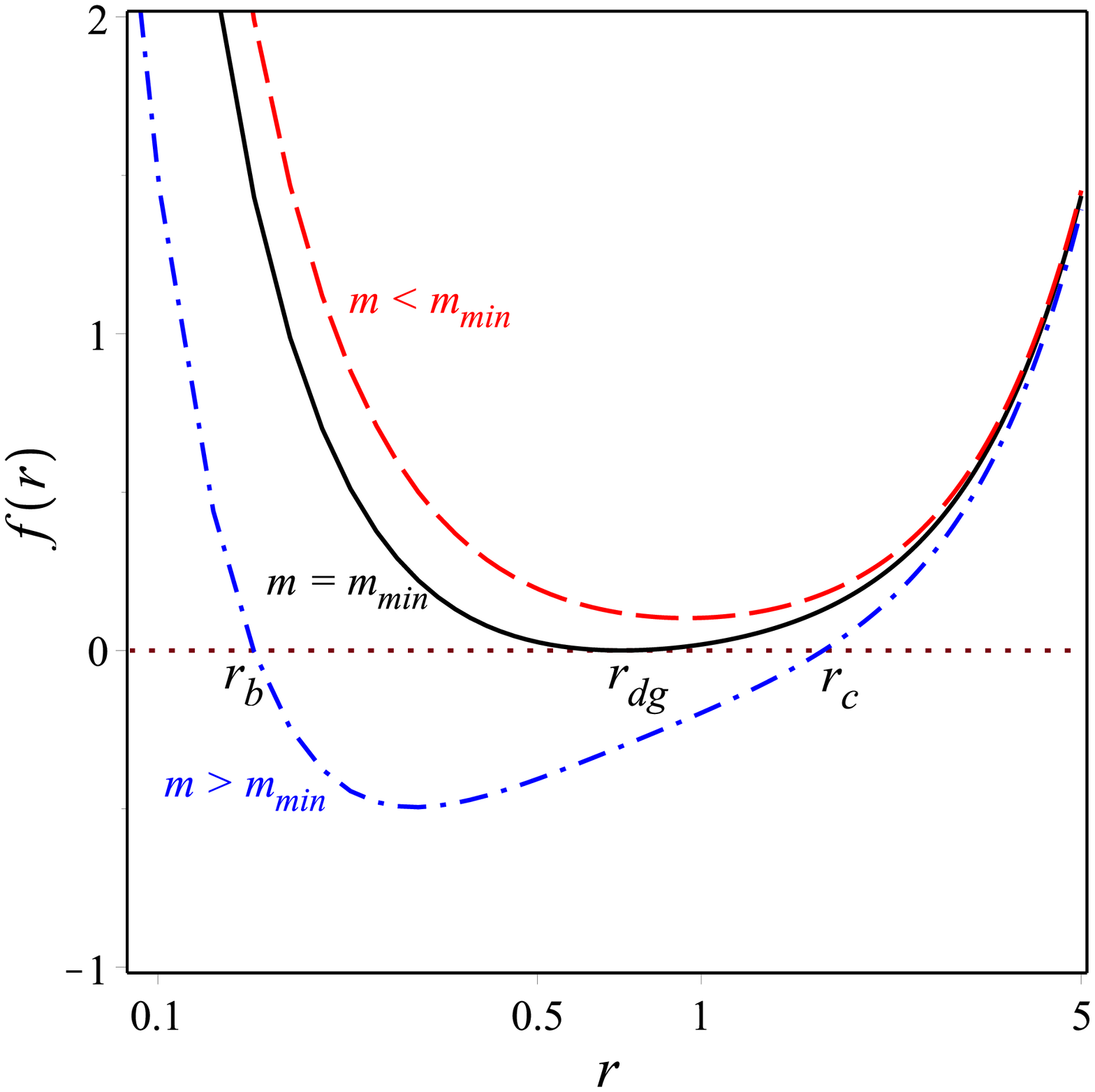}}\hspace{1cm}
\subfigure[~The degenerate horizon--charge correlation]{\label{fig:2b}\includegraphics[scale=0.35]{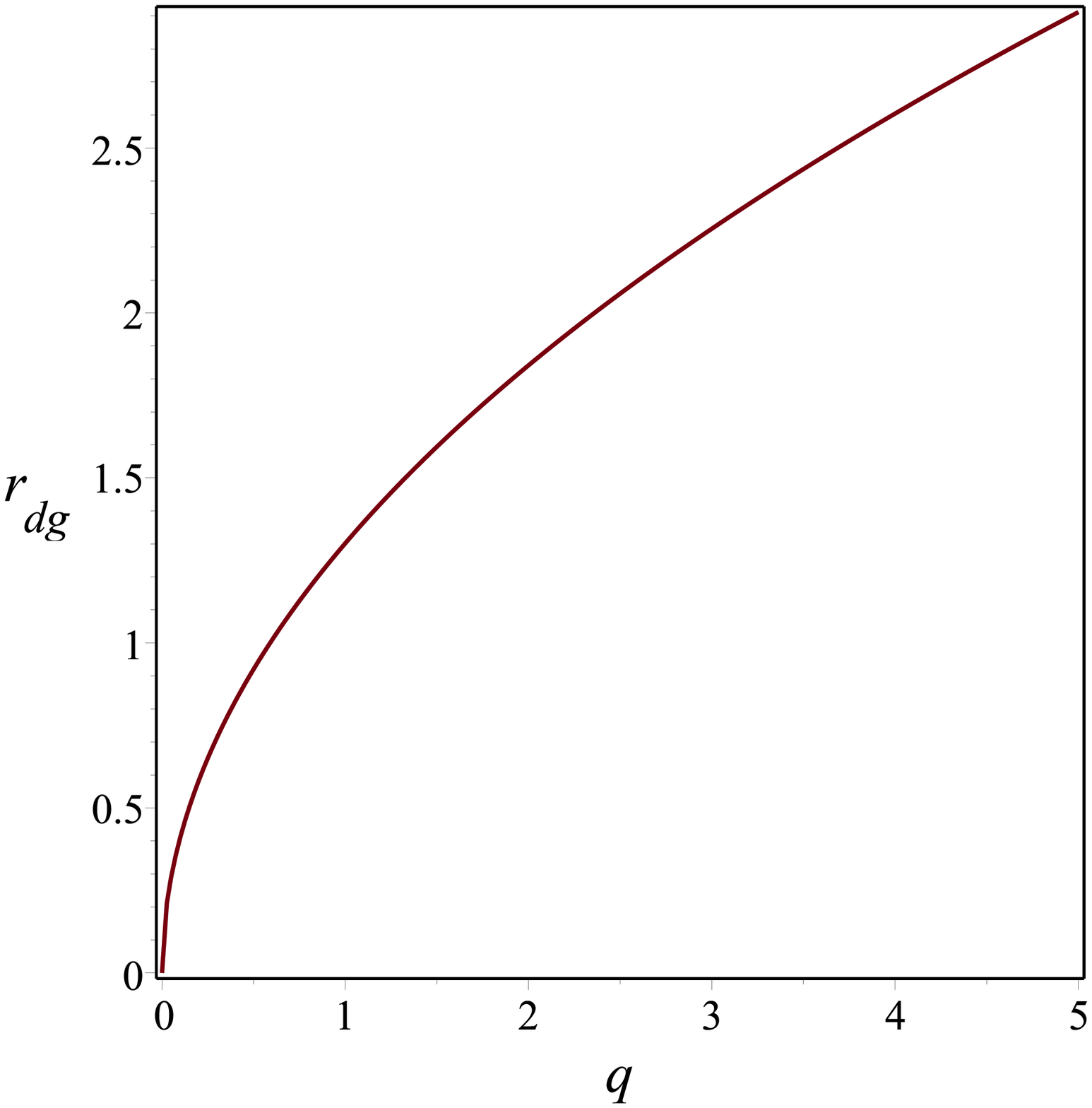}}
\caption{Schematic plots of the degenerate horizons in the linear Maxwell mimetic gravity: \subref{fig:2a} The plot of $f(r)$ shows the black hole event horizon, $r_b$, and the cosmological horizon, $r_c$, where $m>m_{min}$. At $m=m_{min}$ ($r_b=r_c$), the black hole has a degenerate horizon at $r_{dg}$. Otherwise, $m<m_{min}$ the black hole is naked; \subref{fig:2b} The plot shows the dependence of the degenerate horizon on the charge $q$ at fixed mass $m=m_{min}$. Here we take $D=4$ and $\Lambda=0.058$.}
\label{Fig:2}
\end{figure*}

One can calculate the horizons of the linear, Eq. (\ref{m11}), and the nonlinear, Eq. (\ref{m22}), electrodynamics cases by finding the roots of   $f(r) = 0$. In $4$-dimensional spacetime, these can be seen in Figs. \ref{Fig:1}\subref{fig:1a} and \ref{Fig:1}\subref{fig:1b} for particular values of the model parameters. The plots show the two roots of $f(r)$ which determine the black hole $r_b$  and the cosmological $r_c$ horizons of the solutions at hand  which in agreement with the results of Sec. \ref{S6}, . We note that, in the linear case, for $m>0$, $q>0$ and $\Lambda>0$, we find that these two roots are possible when $m> m_{min}=\frac{2}{3}6^{1/4}q^{3/2}\Lambda^{1/4}$. Interestingly, when $m = m_{min}$, we determine the degenerate horizons $r_{dg}= (q/\sqrt{6\Lambda})^{1/2}$ at which $r_b=r_c$, that is Nariai black hole. The thermodynamics of Nariai black hole has been studied in several works, c.f. \cite{Myung:2007qt,Kim:2008hm,Myung:2007av}. Otherwise, when $m<m_{min}$, there is no black hole. This can be shown in Fig. \ref{Fig:2}\subref{fig:2a}, while Fig. \ref{Fig:2}\subref{fig:2b} is showing the dependence of the degenerate horizon on the charge in the linear electrodynamics case. Similarly, the degenerate horizon can be determined in the nonlinear electrodynamics case. In $4$-dimensional spacetime, the solutions which have two horizons similar to our model can be obtained for Schwarzschild-dS and Kerr-Schild class \cite{Di96,Dymnikova:2001fb, Dymnikova:2018uyo}, RN black holes surrounded with quintessence \cite{Ghaderi:2016dpi}, minimal model of a regular black hole \cite{Hayward:2005gi}, and also in the case of Bardeen black holes which are spherically symmetric solution of the noncommutative geometry \cite{Kim:2008hm,Myung:2007av, Nicolini:2005vd,Sharif:2011ja}. In our calculations, we use positive values of the cosmological constant, since we are interested in the double horizons solutions. However, it is worth to mention that negative cosmological constant produces a pattern similar to the case of two vacuum scales spacetime which connects two de Sitter vacua at $r\to 0$ and $r\to \infty$. The later is characterized by at most three horizons \cite{Bronnikov:2003yi,Bronnikov:2012mf}. However, in our case we have exactly one horizon.


The black hole thermodynamical stability is related to the sign of its heat capacity $C_h$. In the following, we analyze the thermal stability of the black hole solutions via the behaviour of their heat capacities \cite{Nouicer:2007pu,DK11,Chamblin:1999tk}
\begin{equation}\label{m55}
C_h=\frac{dE_h}{dT_h}= \frac{\partial m}{\partial r_h} \left(\frac{\partial T}{\partial r_h}\right)^{-1},
\end{equation}
where $E_h$ is the energy.
If the heat capacity $C_{h} > 0$ ($C_h < 0$), the black hole is thermodynamically stable (unstable), respectively. To a better understanding of this process, we assume that at some point and due to thermal fluctuations, the black hole absorbs more radiation than it emits, which means that its heat capacity is positive. This means that the mass of the black hole indefinitely increases. On the contrary, the black hole emits more radiation than it absorbs, which means the heat capacity is negative. This means that the black hole mass indefinitely decreases until it disappears completely. Thus, black holes with negative heat capacities are thermally unstable.

\begin{figure*}
\centering
\subfigure[~Horizon mass--radius relation]{\label{fig:3a}\includegraphics[scale=0.28]{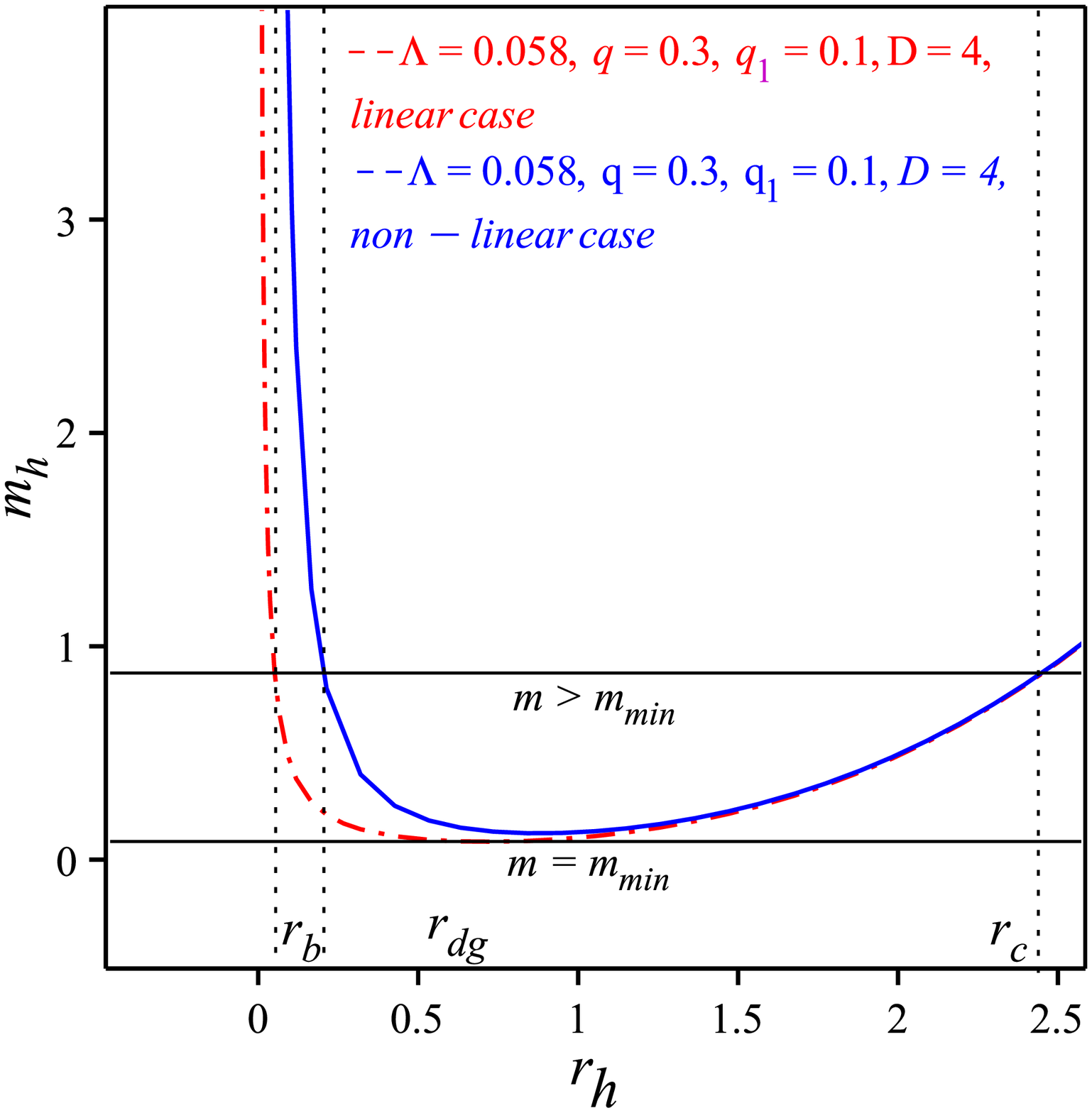}}\hspace{0.2cm}
\subfigure[~Horizon temperature--radius relation]{\label{fig:3b}\includegraphics[scale=0.28]{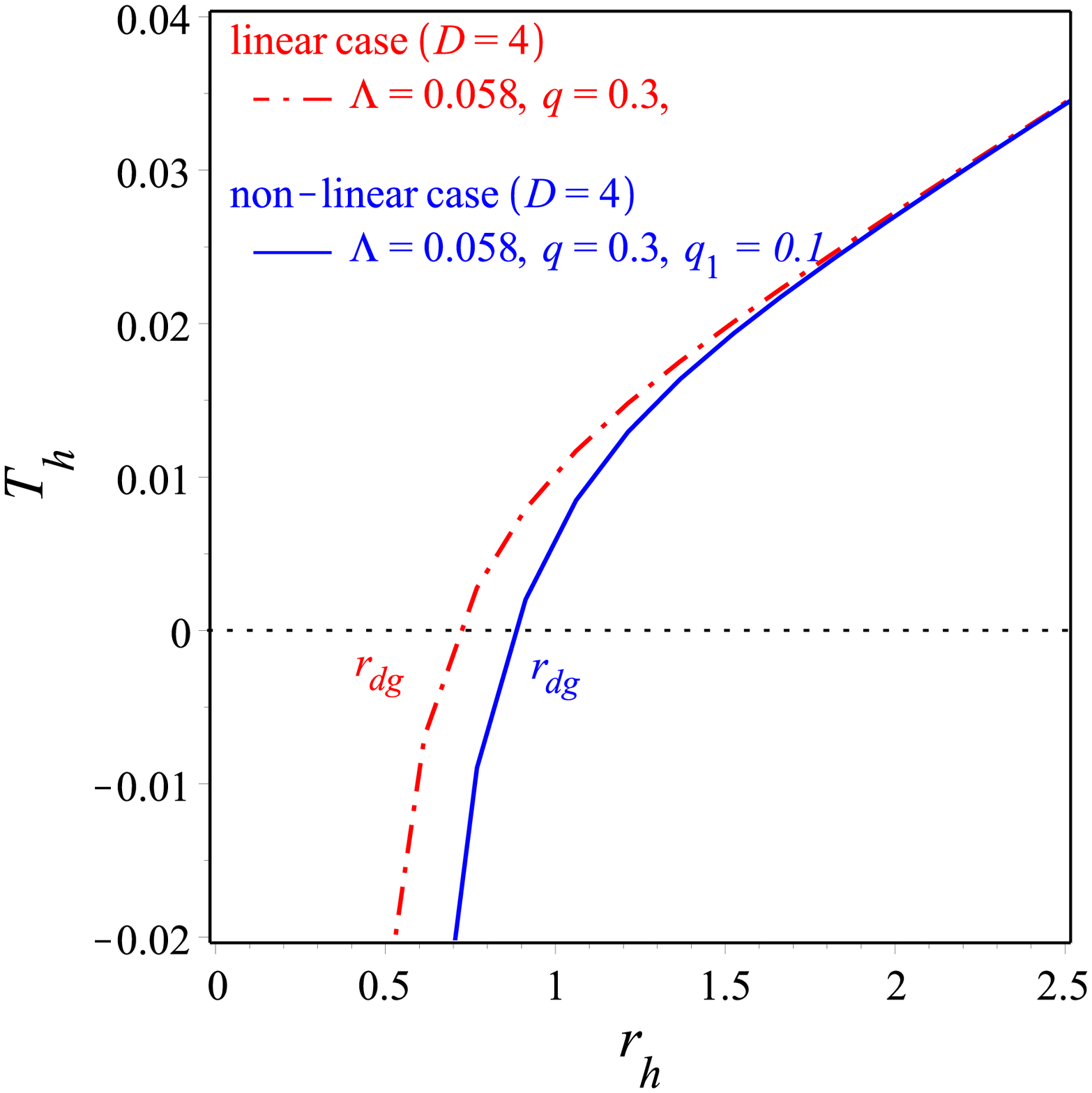}}\hspace{0.2cm}
\subfigure[~Horizon heat capacity--radius relation]{\label{fig:3c}\includegraphics[scale=0.28]{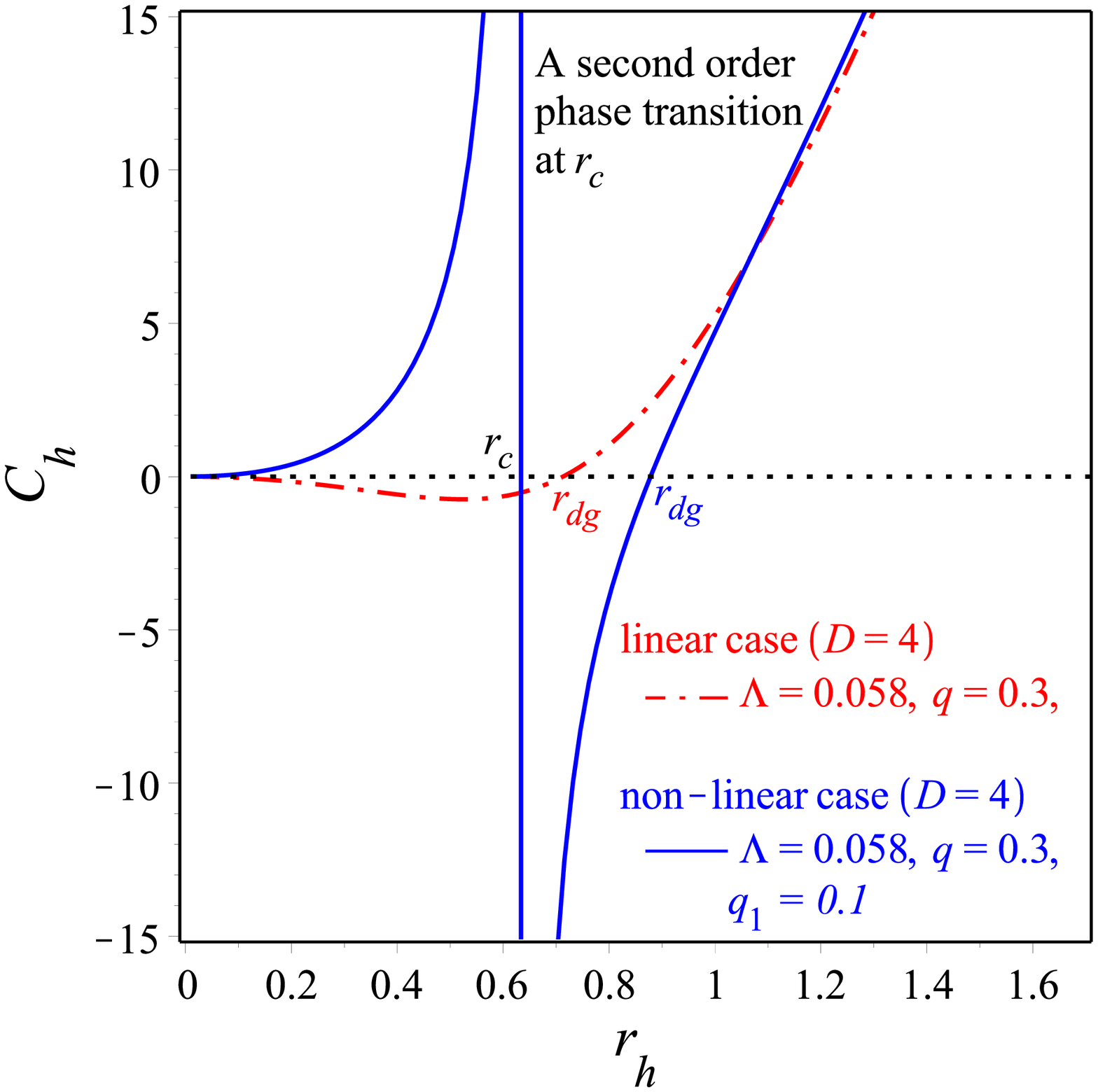}}
\caption{Schematic plots of some thermodynamical quantities of the black hole solutions: \subref{fig:3a} The horizon mass-radius relation determines the minimal mass required to identify the degenerate horizon; \subref{fig:3b} The temperature vanishes on the horizon radius $r_h$; \subref{fig:3c} The heat capacity shows the locally unstable event horizon which is characterized by the negative $C_h<0$. Also, the nonlinear electrodynamics case shows a second-order phase transition as $C_h$ diverges.}
\label{Fig:3}
\end{figure*}

In order to evaluate Eq. (\ref{m55}), it requires us to derive analytical formulae of $m_h\equiv m(r_h)$ and $T_h \equiv T(r_h)$. Firstly, we calculate the black hole mass within an even horizon $r_h$. We set $f(r_h) = 0$, then we get
\begin{eqnarray} \label{m33}
&&{m_h}_{{}_{{}_{{}_{{}_{\tiny Eq. (\ref{m11})}}}}}=r_h^{(D-3)}\left(r_h^2\Lambda_{eff}+\frac{(D-3)q^2}{(D-2)r_h^{2(D-3)}}\right), \nonumber\\
&&{m_h}_{{}_{{}_{{}_{{}_{\tiny Eq. (\ref{m22})}}}}}=r_h^{(D-3)}\left(r_h^2\Lambda_{eff}+\frac{q^2}{(D-2)(D-3)r_h^{2(D-3)}}+\frac{q_1}{(D-2)^2r_h^{2D-5}}+\frac{q_1^2}
{4(D-2)(2D-5)r_h^{(3D-8)}}\right).
\end{eqnarray}
The above equations show that the black hole total mass is given as a function of the horizon radius and the charge. One can also find the degenerate horizon by setting $\partial m_h/\partial r_h=0$, in $4$ dimensions; this gives $r_{dg}=(q/\sqrt{6\Lambda})^{1/2}$ as previously obtained. For fixed charge values in both linear and nonlinear charged black holes, we plot the horizon mass-radius relation as is depicted in Fig. \ref{Fig:3}\subref{fig:3a}. As seen from this figure, the horizon mass--radius relation is characterized by
\begin{equation} \label{m333}
m(r_h\rightarrow 0)\rightarrow \infty, \qquad \qquad m(r_h\rightarrow \infty)\rightarrow \infty.\end{equation}
Also, one can find that there is a minimal mass at the degenerate horizon whereas the double horizons (event and cosmological) coincide. For larger masses, the double horizons are separated, while smaller masses do not show horizons. This confirms the results of Fig. \ref{Fig:2}\subref{fig:2a}. In this sense, we find that the model at hand shares some features with the minimal model of a regular black hole \cite{Hayward:2005gi}. However, as shown by Eq. (\ref{m333}), there is no minimal length of the black hole event horizon as in the minimal model scenario.


The Hawking temperature of the black holes can be obtained by requiring the absence of  singularity at the horizon in the Euclidean sector of the black hole solutions. Secondly, we obtain  the associated temperature with the outer event horizon $r = r_h$ as \cite{Hawking:1974sw}
\begin{equation}
T = \frac{\kappa}{2\pi}, \qquad \textmd{where} \quad \kappa \quad \textmd{is the surface gravity defined as } \qquad \kappa= \frac{f'(r_h)}{2}.
\end{equation}
The Hawking temperatures associated with the black hole solutions (\ref{m11}) and (\ref{m22}) are
\begin{eqnarray} \label{m44}
{T_h}_{{}_{{}_{{}_{{}_{\tiny Eq. (\ref{m11})}}}}}&=&\frac{1}{4\pi}\left(\frac{(D-1)(D-2)r_h^{2(D-2)} \Lambda_{eff}-q^2}{(D-2)r_h^{2D-5}}\right), \nonumber\\
{T_h}_{{}_{{}_{{}_{{}_{\tiny Eq. (\ref{m22})}}}}}&=&\frac{1}{4\pi}\Bigg\{\frac{(D-1)(D-2)r_h^{2(D-2)} \Lambda_{eff}-q^2}{(D-2)r_h^{2D-5}}
 +\frac{(D-3)q_1}{(D-2)^2r_h^{2(D-2)}}+\frac{(D-3)q_1^2}
 {4(D-2)(2D-5)r_h^{(3D-7)}}\Bigg\}\nonumber\\
\end{eqnarray}
where ${T_h}$ is the Hawking temperature at the event horizon. For linear and nonlinear electrodynamics, in the $4$-dimensional spacetime, we plot the horizon temperatures--radius relation in Fig. \ref{Fig:3}\subref{fig:3b} for particular values the black hole parameters. The figure shows that the horizon temperature $T_h$ vanishes at the degenerate horizon $r_h=r_{dg}$. At $r_h< r_{dg}$, the horizon temperature goes below absolute zero forming an ultracold black hole. As noted earlier by Davies \cite{Davies:1978mf} that there is no obvious reasons from thermodynamics prevent a black hole temperature to go below absolute zero and turn it to naked singularity. In fact this is the case presented in Fig. \ref{Fig:3}\subref{fig:3b} at $r_h<r_{min}$ region. However, this case of ultracold black hole is justified in the presence of phantom energy field \cite{Babichev:2014lda}, and explains the decreasing mass pattern in Fig. \ref{Fig:3}\subref{fig:3a}. Indeed, this case suits well with mimetic gravity theories whereas the SEC breaking indicate a phantom mimetic field. Also, it would be useful to examine some potential patterns in that case. At $r_h > r_{dg}$, the horizon temperature becomes positive. For values $r_h$ larger enough, the horizon temperatures of both linear and nonlinear charged black holes behave similarly. Including the gravitational effect of thermal radiation, one can show that at some very high temperature $T_{max}$ the radiation would become unstable and collapse to a black hole \cite{Hawking:1982dh}. Hence, the pure AdS solution is only stable at temperatures $T < T_{max}$. Above $T_{max}$, only the heavy black holes would have stable configurations \cite{Hawking:1982dh}.

Next we evaluate the horizon heat capacity $C_h$, therefore we substitute Eqs. (\ref{m33}) and (\ref{m44}) into Eq. (\ref{m55}). Hence, we write
\begin{eqnarray} \label{m66}
&&{C_h}_{{}_{{}_{{}_{{}_{\tiny Eq. (\ref{m11})}}}}}=\frac{2\pi r_h{}^2\left[(D-1)(D-2)r_h{}^{3D-4} \Lambda_{eff}-q^2r_h{}^D\right]}{(D-1)(D-2)r_h{}^{2D} \Lambda +(D-5)r_h{}^4q^2}, \nonumber\\
 &&{C_h}_{{}_{{}_{{}_{{}_{\tiny Eq. (\ref{m22})}}}}}=\frac{(2D-5)\pi r_h{}^{D-2}\left[\left\{[4(D-1)(D-2)r_h{}^{3(D-2)} \Lambda_{eff}-q_1{}^2]r_h{}^{2D-5}-4q_1r_h{}^{3D-8}\right\}r_h{}^{2(D-3)}-4q^2r_h{}^{5D-13}\right]}{r_h{}^{7D-19}\left[(D-5)[4(D-1)(D-2)r_h{}^2 \Lambda +4(2D-5)q^2r_h{}^{-2(D-3)}-8(D-3)r_h{}^{-2(D-3)}q_1]-(3D-7)(D-3)r_h{}^{-(3D-8)}q_1{}^2\right]}.\nonumber\\
 &&\end{eqnarray}
It is not easy to extract information directly from Eqs. (\ref{m66}), therefore we plot them in Fig. \ref{Fig:3}\subref{fig:3c} for particular values of the black hole parameters. As we show on this figure, in both linear and nonlinear charged black holes, the horizon heat capacities vanish at the degenerate horizon $r_{dg}$ as well as their horizon temperatures. In the linear electrodynamics, the heat capacity has negative values for $r_h<r_{dg}$, but it has positive values for $r_h > r_{dg}$. In the nonlinear electrodynamics, the heat capacity is positive for all values of the horizon radius $r_h$ except for an intermediate region $r_{c}< r_h <r_{dg}$, it is negative and characterized by a second-order phase transition at $r_{c}$ whereas the heat capacity has an infinite discontinuity.

In conclusion, in the linear charged black hole case, a typical pattern of the heat capacity has been obtained in literature c.f \cite{Ma:2015llh}. However, we find that the negative heat capacity region is associated with a positive temperature on the contrary to our case. In mimetic gravity with linear electrodynamics, the negative heat capacity region, i.e $r_h<r_{dg}$, is associated with a temperature below absolute zero. This can be justified in the presence of a phantom field as in our case whereas the SEC is not fulfilled, see Eqs. (\ref{cal}). At $r_h=r_{dg}$, the temperature on the event horizon is exactly zero as well as the heat capacity as shown on Figs \ref{Fig:3}\subref{fig:3b} and \ref{Fig:3}\subref{fig:3c}. For the case $r_h>r_{dg}$, both temperature and heat capacity are positive and the solution is in a thermal equilibrium. Indeed, thermodynamics stability of the charged black hole in AdS spacetime has been widely studied in many theories, thermodynamics of Bardeen (regular) black holes c.f.~\cite{Myung:2007qt}, Schwarzschild-AdS in two vacuum scales case \cite{Dymnikova:2010zz}, and also similar work in the noncommutative geometry \cite{Man:2013hpa,Berej:2006cc,Tharanath:2014naa,Maluf:2018lyu}. Indeed, all these solutions are characterized by a second-order phase transition where the heat capacity has an infinite discontinuity similar to our case, while the heat capacity remains negative at $r_h> r_{dg}$. This is in contrast to our case, as clear from Fig. \ref{Fig:3}\subref{fig:3c}, whereas the heat capacity crosses to positive regions as $r_h$ goes to larger values. This qualitative difference is due to the nonlinear electrodynamics contribution. As we have mentioned before, the nonlinear contribution of the model at hand is derived from the equations of motions in contrast to other models which pre-assume the form of the nonlinear terms to produce RN asymptotically.
\section{Summary and Prospectives}\label{S8}
Day after day dark matter is being confirmed by astrophysical and cosmological observations. Two main streams have been proposed to explain dark matter, modifications of Einstein's general relativity and modifications of the standard model by introducing new particle species. It has been shown that these two classes are not different after all \cite{Calmet:2017voc}. In fact, every modified gravity model has new degrees of freedom besides the usual massless graviton. As we mentioned in the introduction that the mimetic gravity is a good candidate to explain CDM presence. This motivates us to explore the theory in astrophysics domain, by investigating possible new solutions of charged rotating black holes.

For this aim, we derive the field equations of Maxwell mimetic gravitational theory. We apply these field equations to $D$-manifold with $\ell$ angular coordinates, $k$-dimension Euclidean metric and one unknown function $f(r)$ of the radial coordinate. in addition, we use a generalized vector potential which includes three unknown functions, one of them is related to the electric charge and the other two functions are related to the magnetic field. In this context, we derive charged $D$-dimension black hole solutions that possess the mass and the electric charge of the black hole. These behave asymptotically as (A)dS. Then, we apply a coordinate transformation relating the angular coordinates and the temporal coordinate, which allows to derive $D$-dimension rotating charged black hole solutions.

Similarly, in the nonlinear electrodynamics coupled to mimetic gravity, we derive $D$-dimension charged black hole solutions. We apply the corresponding field equations which allow to obtain a new $D$-dimension charged black hole solutions. Interestingly, the new black holes show interesting physical properties, besides the monopole term the solutions contain nonlinear effects related to the dipole and quadrupole charges. The later terms are characterized by a common constant, so that its vanishing derives the solution to the linear case. We show that the asymptotic behaviours of this class of the black hole solutions behaves as (A)dS spacetime. Similar to the Maxwell electrodynamics case, we apply a coordinate transformation which allows to obtain new $D$-dimension rotating charged black hole solutions analytically.

Also, we calculate the invariants constructed from the curvature, namely Kretschmann invariants $K$, the Ricci tensors squared $R_{\mu \nu}R^{\mu \nu}$ and the Ricci scalars $R$, to study possible singularities of the obtained solutions. This shows that they have a true singularities at $r=0$. However, in the Maxwell mimetic gravity, the asymptotic behaviour of these invariants are $K=R_{\mu \nu}R^{\mu \nu} \sim r^{-4(D-2)}$ and $R \sim  r^{-2(D-2)}$. On the other hand, their asymptotic behaviours are $K=R_{\mu \nu}R^{\mu \nu} \sim r^{-6(D-2)}$ and $R \sim  r^{-3(D-2)}$. This indicates that the nonlinear electrodynamics mimetic gravity produces singularities stronger than the Maxwell electrodynamics case. In addition, we calculate the number of horizons, in the linear (nonlinear) case the solutions have $D-2$ horizons. This again proves that the singularities of the black hole solutions in the nonlinear case are stronger than the linear case. Then, we investigate the fulfilment of the energy conditions in the context of the obtained solutions. We verified the fulfillment of the WEC which guarantees the positivity of the energy density. Supported by the preliminary studies \cite{Barvinsky:2013mea,Chaichian:2014qba}, the obtained results are in favor of the absence of ghosts. However, we remind that the perturbation analysis of the present theory is a mandatory to fully investigate the ghost problem. Since here we focus on finding new solutions of charged black holes, we leave the perturbation analysis to be carried out separately elsewhere in the future.

Finally, for the linear and the nonlinear electrodynamics cases, we study the horizons showing that the solutions could have at most two horizons, a black hole event horizon $r_b$ and a cosmological one $r_c$. We determine a minimum value of the black hole mass at which the two horizons coincide forming the degenerate horizon, above this minimum mass the black hole would have two horizons, below the minimum mass it there is no black hole. Also, we study thermodynamics and thermal phase transitions of the obtained black holes.  In the linear electrodynamics case, the temperature drops below absolute zero when $r_h<r_{min}$ forming an ultracold black hole, while the heat capacity is being negative at these regions and so it is unstable. Similar conclusions can be derived for the nonlinear electrodynamics case, however it is characterized by a second-order phase transition whereas the heat capacity has an infinite discontinuity.

\subsection*{Acknowledgments}
The work of GGLN is partially supported by the Egyptian Ministry of Scientific Research under project No. 24-2-12. Moreover, the work of KB was supported in part by the JSPS KAKENHI Grant Number JP 25800136 and Competitive Research Funds for Fukushima University Faculty (18RI009).

%

\end{document}